\documentclass[twocolumn,polish,english,aps,prl,superscriptaddress,showpacs]{revtex4}
\usepackage[T1]{fontenc}
\usepackage[latin2,latin9]{inputenc}
\usepackage{amstext}
\usepackage{amssymb}
\usepackage{esint}

\makeatletter
\@ifundefined{textcolor}{}
{%
 \definecolor{BLACK}{gray}{0}
 \definecolor{WHITE}{gray}{1}
 \definecolor{RED}{rgb}{1,0,0}
 \definecolor{GREEN}{rgb}{0,1,0}
 \definecolor{BLUE}{rgb}{0,0,1}
 \definecolor{CYAN}{cmyk}{1,0,0,0}
 \definecolor{MAGENTA}{cmyk}{0,1,0,0}
 \definecolor{YELLOW}{cmyk}{0,0,1,0}
}


\makeatother

\usepackage{babel}
\begin{document}

\title{Occam's Razor Cuts Away the Maximum Entropy Principle}

\author{\L{}ukasz Rudnicki}

\email{rudnicki@cft.edu.pl}

\affiliation{Freiburg Institute for Advanced Studies, Albert-Ludwigs University
of Freiburg, Albertstrasse 19, 79104 Freiburg, Germany}

\affiliation{Center for Theoretical Physics, Polish Academy of Sciences, Aleja
Lotnik\'ow 32/46, PL-02-668 Warsaw, Poland}
\begin{abstract}
I show that the maximum entropy principle can be replaced by a more
natural assumption, that there exists a phenomenological function
of entropy consistent with the microscopic model. The requirement
of existence provides then a unique construction of the related probability
density. I conclude the letter with an axiomatic formulation of the
notion of entropy, which is suitable for exploration of the non-equilibrium
phenomena. 
\end{abstract}
\maketitle
The maximum entropy principle entered physics as a conclusion drawn
by Gibbs from his description of classical statistical mechanics \cite{Gibbs}.
In its most proper form it was established by the theorem saying that:
``\emph{If an ensemble of systems is canonically distributed in phase,
the average index of probability is less than in any other distribution
of the ensemble having the same average energy}'' %
\footnote{In \cite{Gibbs} this statement can be found as \emph{Theorem II}
present on page 130.%
}. Using a bit more modern language ``the average index of probability''
is the same as the average logarithm of the density, while the word
``phase'' simply refers to the phase space. I use the term ``density'',
in order to simultaneously cover both cases of a classical probability
distribution and a quantum density operator. After Shannon had promoted
the entropy to be the major quantity in information theory \cite{Shannon},
the maximum entropy principle became, due to Jaynes, one of the fundamental
laws of physics \cite{Jaynes1,Jaynes2,Werhl}. During the next more
than 50 years this principle has found hundreds of applications in
statistical mechanics (with emphasis on non--equilibrium phenomena)
and information theory \cite{RMP,Ingarden}. With relatively little
effort one can find examples from before the Jaynes formulation, in
which the maximum entropy principle also played an important role
in development of new theoretical concepts, such as relativistic thermodynamics
\cite{Relativistic}. 

It is probably a common feeling that the nature's tendency to maximize
the entropy possesses a deeper philosophical meaning \cite{Uffink1,Uffink2}.
While many scientists accept this tendency as being typical for physical
theories, the unavoidable effort necessary to pick up the ``maximal''
scenario might raise some doubts (especially when it concerns the
theory aiming at quantifying all kinds of efforts). The effort in
question splits in fact into two subsequent tasks. First of all, we
must exclude all cases which make the entropy depend on more average
quantities (like the average energy) than anticipated. But this step
we can as well make on the phenomenological level, by assuming that
the entropy does depend only on the variables we are to use (or can
experimentally access). There is no true necessity to invoke constrained
optimization, to get rid of information we do not have anyway. It
is enough to say that we fully rely on the information which is accessible
for us. The second task is the following: there exist many densities
providing the entropy as a function of wanted parameters only, so
it becomes necessary to select the maximal option. 

So, is it eventually possible to avoid the requirement that the entropy
must be maximal? The answer is yes, provided that the second problem
listed above can be solved in a much simpler and physically more natural
way. The aim of this letter is thus to prove that for a fixed, finite
number of average parameters there is always\emph{ only} one density
$\varrho_{0}$ such that the entropy
\begin{equation}
-k_{B}\left\langle \ln\varrho\right\rangle _{\varrho},\label{entropy}
\end{equation}
evaluated\textbf{ }for\textbf{ }$\varrho=\varrho_{0}$ depends \emph{only}
(!) on these parameters, and the form of such \emph{phenomenological
entropy function} $\mathcal{S}$ is preserved by all infinitesimal
fluctuations of $\varrho_{0}$. According to a common notation $\left\langle \cdot\right\rangle _{\varrho}$
denotes the average with respect to $\varrho$, while $k_{B}$ is
the Boltzmann constant but can as well be an arbitrary constant with
a proper unit. 

Before going into the details let me once more state the main message
of this letter. Assume that we restrict ourselves to the phenomenological
description based on a finite number of average parameters supplemented
by the parameters which are constant (like the volume and the number
of particles in the canonical ensemble). There exists the unique choice
of the density (naturally the same as obtained by maximization \cite{Jaynes1,Jaynes2})
such that the entropy function depends only on the selected parameters
and is microscopically given by the formula (\ref{entropy}). The
above statement happens to be too strong to be valid in general, since
any family of densities involving a proper number of parameters can
eventually be a good candidate for $\varrho_{0}$. But to make it
true, it is sufficient to assume that whenever we infinitesimally
change the density $\varrho_{0}$ by $\delta\varrho$, the form of
the phenomenological entropy $\mathcal{S}$ remains the same, while
the values of the involved average parameters change accordingly. 

We can thus convert the maximum entropy principle to be the more plausible
requirement of existence. As I show in the latter part, this remarkable
property supports the microscopic definition of the entropy (\ref{entropy}),
because other choices do not necessarily assure $\varrho_{0}$ to
be uniquely defined. 

\paragraph{The main result.---}

Let me start the main discussion of this letter with few remarks about
the notation. Considering a general landscape it becomes necessary
to distinguish two sets of parameters. First of all we chose the variables
$\left\{ \mathcal{V}\right\} =\left\{ \mathcal{V}_{1},\mathcal{V}_{2}\ldots\right\} $
which are assumed to be\emph{ externally fixed} (the number of these
variables does not need to be specified). All variables relevant to
the microcanonical ensemble (energy, volume, number of particles)
do belong to this set, while in the case of the canonical ensemble
only the volume and the number of particles remain externally fixed.
The second set of parameters is crucial for the Jaynes formulation
of the maximum entropy principle \cite{Jaynes1,Jaynes2}. It consists
of $M$ additional variables ($j=1,\ldots,M$) 
\begin{equation}
F_{j}\equiv F_{j}\left[\varrho_{0}\right],\label{SecConstr}
\end{equation}
represented by the linear functional $F_{j}\left[\varrho\right]=\left\langle \tilde{F}_{j}\right\rangle _{\varrho}$
given in terms of the average values of some properly selected quantities
$\tilde{F}_{j}$. In (i) the classical case, $\tilde{F}_{j}$ are
simply functions of the phase-space variables, while in (ii) the quantum
case they are all Hermitian operators $\hat{F}_{j}$. We shall further
distinguish the simplified case (ii-a) when all the operators commute
with each other and the general case (ii-b) involving possibly non-commuting
quantities. 

Since the maximum entropy principle relies on constrained optimization,
let me make a conceptual distinction. The naturally (always) present
constraint on the norm of the distribution, $\left\langle 1\right\rangle _{\varrho}=1$,
I shall call the \emph{primary constraint}, while the additional constraints
I shall call secondary. According to the maximum entropy principle,
the density $\varrho_{0}$ defining $F_{j}$ is such that the entropy
(\ref{entropy}) becomes maximal, provided that the primary constraint
and the secondary constrains (\ref{SecConstr}) are satisfied. In
this letter I shall provide an alternative derivation of the proper
density $\varrho_{0}$.

While we do not plan to use the secondary constraints (\ref{SecConstr})
directly, we cannot get rid of the primary constraint. We shall thus
incorporate it into the analysis by defining the \emph{microscopic
entropy functional} \inputencoding{latin2}\foreignlanguage{polish}{
\begin{equation}
S_{\textrm{m}}\left[\varrho\right]=-k_{B}\left\langle \ln\varrho\right\rangle _{\varrho}-\lambda\left(\left\langle 1\right\rangle _{\varrho}-1\right).\label{zmodyfikowana}
\end{equation}
}\inputencoding{latin9}Since $\lambda$ is an arbitrary parameter,
the above formula strongly resembles the method of Lagrange multipliers.
The term $\lambda\left(\left\langle 1\right\rangle _{\varrho}-1\right)$
is however here rather an artificial gauge which does not change the
value of entropy, as long as the density $\varrho$ is normalized.
Even though Eq. (\ref{zmodyfikowana}) prepares the reader for an
optimization routine and the next steps can bring more such similarities,
I would strongly like to emphasize that the whole reasoning has nothing
to do with any kind of optimization. All the formulas appearing
below, being indeed very similar to those in the method of the Lagrange
multipliers, are purely functional identities.

Let me now assume that the phenomenological entropy 
\begin{equation}
\mathcal{S}=\mathcal{S}\left(F_{1},\ldots,F_{M};\left\{ \mathcal{V}\right\} \right),\label{macros}
\end{equation}
 is a function of the parameters $\left\{ F\right\} =F_{1},\ldots,F_{M}$
and $\left\{ \mathcal{V}\right\} $. Let me further introduce the
corresponding \emph{phenomenological entropy functional}\inputencoding{latin2}\foreignlanguage{polish}{
\begin{equation}
S_{\textrm{ph}}\left[\varrho\right]=\mathcal{S}\left(F_{1}\left[\varrho\right],\ldots,F_{M}\left[\varrho\right];\left\{ \mathcal{V}\right\} \right).\label{phen}
\end{equation}
}\inputencoding{latin9}By construction, we have that $\mathcal{S}\equiv S_{\textrm{ph}}\left[\varrho_{0}\right]$,
so the macroscopic (phenomenological) entropy function $\mathcal{S}$
is given by the functional (\ref{phen}) evaluated for $\varrho=\varrho_{0}$.

Note that the functionals $S_{\textrm{m}}\left[\varrho\right]$ and
$S_{\textrm{ph}}\left[\varrho\right]$ are defined on the whole domain
of $\varrho$ and there are no secondary constraints spoiling this
property. We can thus easily calculate the functional derivatives
of both functionals:
\begin{equation}
\frac{\delta S_{\textrm{m}}\left[\varrho\right]}{\delta\varrho}=-k_{B}\left(1+\ln\varrho\right)-\lambda,\label{fd1}
\end{equation}
\begin{equation}
\frac{\delta S_{\textrm{ph}}\left[\varrho\right]}{\delta\varrho}=\sum_{j=1}^{M}\left(\frac{\partial\mathcal{S}}{\partial F_{j}}\right)_{F_{k\neq j}}\frac{\delta F_{j}\left[\varrho\right]}{\delta\varrho},\label{fd2}
\end{equation}
with the quantity $\delta F_{j}\left[\varrho\right]/\delta\varrho$
being equal to $\tilde{F}_{j}$. The derivatives $\left(\partial\mathcal{S}/\partial F_{j}\right)_{F_{k\neq j}}$
are $\varrho$-dependent functionals, and the thermodynamic notation
$\left(\cdot\right)_{F_{k\neq j}}$ has a usual meaning that we differentiate
with respect to $F_{j}$ keeping constant all other variables $F_{k}$,
for $k\neq j$.

We are now ready to formally establish the main result of this letter.
For a given set of parameters $\left\{ F\right\} $ and $\left\{ \mathcal{V}\right\} $,
there exist a unique density $\varrho_{0}$ and a unique phenomenological
entropy function $\mathcal{S}$, such that
\begin{equation}
S_{\textrm{m}}\left[\varrho_{0}+\delta\varrho\right]-S_{\textrm{ph}}\left[\varrho_{0}+\delta\varrho\right]=\mathcal{O}\left(\left(\delta\varrho\right)^{2}\right),
\end{equation}
or equivalently:
\begin{equation}
S_{\textrm{m}}\left[\varrho_{0}\right]=S_{\textrm{ph}}\left[\varrho_{0}\right]\equiv\mathcal{S},\quad\left.\frac{\delta S_{\textrm{m}}\left[\varrho\right]}{\delta\varrho}\right|_{\varrho=\varrho_{0}}\!\!\!\!\!\!=\left.\frac{\delta S_{\textrm{ph}}\left[\varrho\right]}{\delta\varrho}\right|_{\varrho=\varrho_{0}}.\label{main2}
\end{equation}
The physical meaning of the above conditions is straightforward. The
left equation in (\ref{main2}) tells us that if $\varrho=\varrho_{0}$,
the macroscopic entropy $\mathcal{S}$ is not only given by the phenomenological
entropy functional (what is true \emph{per se}), but simultaneously
originates from the microscopic model. It is however possible to find
infinitely many couples of densities and entropy functions satisfying
this matching requirement. The second, right condition is especially
interesting. For a given couple $\left(\varrho_{0},\mathcal{S}\right)$
fulfilling the left condition, we scan the infinitesimal neighborhood
of $\varrho_{0}$ and test if the form of the function $\mathcal{S}$
is preserved. We expect that the true phenomenological entropy is
attributed to the particular system treated as a whole, eg. it captures
the nature of the two-body interaction. On the other hand, possible
infinitesimal fluctuations of the density, while enter the microscopic
model, cannot affect the macroscopic character of the system in question
(they cannot lead to a different global interaction mechanism). They
could eventually change the values of the parameters describing the
system, such as the average energy. From the physical perspective,
this \emph{stability} requirement is nothing more than a natural consequence
of the fact, that the phenomenological entropy we have in mind, does
really exist. Once more, let me emphasize that the conditions (\ref{main2})
say nothing about the optimization. They only give a mathematical
meaning to our expectations, we have in relation to the macroscopic
entropy $\mathcal{S}$. 

Using the formulas (\ref{fd1}, \ref{fd2}) we can solve the second
equation from (\ref{main2}) with respect to $\varrho_{0}$, so that
after taking into account the primary constraint we obtain the well--known
expression for the density:
\begin{equation}
\varrho_{0}=\frac{e^{-\sum_{l}\beta_{l}\tilde{F}_{l}}}{\left\langle 1\right\rangle _{e^{-\sum_{m}\beta_{m}\tilde{F}_{m}}}},\qquad\beta_{j}=\frac{1}{k_{B}}\left(\frac{\partial\mathcal{S}}{\partial F_{j}}\right)_{F_{k\neq j}}.\label{density}
\end{equation}
The thermodynamic derivatives defining $\beta_{j}$ are no longer
functionals, but since $\varrho=\varrho_{0}$ they become simple derivatives
of the function (\ref{macros}). The density (\ref{density}) is given
by the exponential solution, similar in form to the solution provided
by the constrained optimization. It is not incredibly surprising,
because the exponential densities are known to be distinguished by
the information--theoretic perspective \cite{entrExp}, and are the
unique distributions possessing a sufficient statistics \cite{suffstat}.
The one and major difference is that in the optimization routine $\beta_{j}$
are the Lagrange multipliers which must be found in such a way that
the entropy becomes maximal. In our current case, these variables
are the inverses of generalized temperatures (derivatives of the entropy). 

It is not true that for any choice of the microscopic entropy functional,
the $\beta_{j}$ parameters would correspond to the derivatives of
$\mathcal{S}$. However for (\ref{zmodyfikowana}) the above consistency
requirement is satisfied, what seems to be a well--known fact in statistical
mechanics \cite{Ingarden2}. Up to now, I have shown that the family
of densities of the same form as given by the maximum entropy principle
can be obtained without resorting to optimization. We could however
expect, that in general it is possible to find many sets of parameters
$\beta_{j}$, such that they are consistent with the secondary constraints
(\ref{SecConstr}) applied \emph{a posteriori}. At that stage the
crucial role of the maximum entropy principle would thus be to pick
up the right set of s $\beta_{j}$. But what if the last problem always
possesses a unique solution? Then the maximum entropy principle can
be completely eliminated in favour of the ``phenomenologically motivated''
condition of existence. The aim of the next paragraph is to prove
that this scenario indeed occurs.

The secondary constraints (\ref{SecConstr}) calculated for the exponential
density (\ref{density}) always provide the relation 
\begin{equation}
F_{j}=f_{j}\left(\beta_{1},\ldots,\beta_{M}\right),\label{system}
\end{equation}
with $f_{j}$ being some functions specific for the particular set
of quantities $\tilde{F}_{j}$. If we assume that the number of average
quantities $M$ is finite, then the above formula in fact describes
a map from $\mathbb{R}^{M}$ to itself. In order to discuss the number
of possible solutions to the system (\ref{system}), we shall characterize
the invertibility property of that map. This however means that we
need to study its Jacobian matrix $J_{ij}=\partial f_{j}/\partial\beta_{i}$.
In the cases (i) and (ii-a) we can easily find that the Jacobian matrix
is: 
\begin{equation}
J_{ij}=-\left(\left\langle \tilde{F}_{i}\tilde{F}_{j}\right\rangle _{\varrho_{0}}-\left\langle \tilde{F}_{i}\right\rangle _{\varrho_{0}}\left\langle \tilde{F}_{j}\right\rangle _{\varrho_{0}}\right).
\end{equation}
The first term inside the parenthesis comes from the derivative of
$e^{-\sum_{l}\beta_{l}\tilde{F}_{l}}$, while the norm $\left\langle 1\right\rangle _{e^{-\sum_{m}\beta_{m}\tilde{F}_{m}}}$
is responsible for the second one. The Jacobian matrix is equal to
minus the covariance matrix evaluated for the set of quantities $\tilde{F}_{j}$.
We thus obtain a very important conclusion: if the quantities $\tilde{F}_{j}$
are chosen in such a way that they are \emph{linearly independent},
then their covariance matrix is positive-definite, and the map (\ref{system})
is everywhere locally invertible. But if instead of all the variables
$F_{j}$ we consider $-F_{j}$, then the Jacobian matrix of the corresponding
map sending $\left(\beta_{1},\ldots,\beta_{M}\right)$ to $\left(-F_{1},\ldots,-F_{M}\right)$
will be positive-definite as well. This however turns out to be the
sufficient condition for a global invertibility of the map \cite{Inject1,Inject2}
so that there always exists a unique solution $\beta_{l}=\left(f^{-1}\right)_{l}\left(-F_{1},\ldots,-F_{M}\right)$.

The case (ii-b) is much more technical, because in order to evaluate
the derivatives of $e^{-\sum_{l}\beta_{l}\hat{F}_{l}}$ we need to
use the operator formula \cite{Operator} 
\begin{equation}
\partial_{\eta}e^{-\hat{A}\left(\eta\right)}=-\int_{0}^{1}dz\, e^{\left(z-1\right)\hat{A}\left(\eta\right)}\frac{\partial\hat{A}\left(\eta\right)}{\partial\eta}e^{-z\hat{A}\left(\eta\right)}.
\end{equation}
It turns out \cite{Ingarden2,Suppl}, that the Jacobian matrix is
equal to $J_{ij}=-2\int_{0}^{1/2}\!\!\! dz\textrm{CM}\left(z\right)$,
and involves the symmetrized covariance matrix \cite{Guhne} $\textrm{CM}\left(z\right)=$
\begin{equation}
\frac{1}{2}\left\langle \hat{F}_{i}\left(z\right)\hat{F}_{j}^{\dagger}\left(z\right)+\textrm{h.c.}\right\rangle _{\varrho_{0}}-\left\langle \hat{F}_{i}\left(z\right)\right\rangle _{\varrho_{0}}\left\langle \hat{F}_{j}\left(z\right)\right\rangle _{\varrho_{0}},
\end{equation}
evaluated for the set of \emph{dressed}, non-Hermitian operators $\hat{F}_{i}\left(z\right)=e^{-z\hat{R}}\hat{F}_{i}e^{z\hat{R}}$,
with $\hat{R}=-\frac{1}{2}\sum_{l}\beta_{l}\tilde{F}_{l}$. In fact,
only the first term of the above covariance matrix depends on $z$,
because the operators $e^{\pm z\hat{R}}$ cancel each other under
the average of a single dressed operator. A much more important observation
is however that the procedure of dressing does not change the mutual
relations between the operators. As long as $\hat{F}_{i}$ are chosen
independently, their counterparts $\hat{F}_{i}\left(z\right)$ are
also linearly independent. Since for every value of $z$ the symmetric
covariance matrix $\textrm{CM}\left(z\right)$ must be positive-definite
\cite{Robertson}, this property is inherited by $-J_{ij}$. We can
immediately apply the previous reasoning to complete the whole proof.

\paragraph{Discussion.---} The most important conclusion from the
above considerations is the fact that we can formulate a new axiomatic
definition of the notion of entropy. It reads: \emph{there exists
a unique choice of the entropy function $\mathcal{S}$ such that:
}(1)\emph{ on the phenomenological level $\mathcal{S}\equiv\mathcal{S}\left(\left\{ F\right\} ;\mathcal{\left\{ V\right\} }\right)$
depends only on a given collection of externally fixed variables $\mathcal{\left\{ V\right\} }$
and a finite number $M$ of average values $F_{j}\equiv F_{j}\left[\varrho_{0}\right]$,
}(2)\emph{ on the microscopic level $\mathcal{S}\equiv S\left[\varrho_{0}\right]$
is given by the formula $S\left[\varrho\right]=-k_{B}\left\langle \ln\varrho\right\rangle _{\varrho}$
evaluated for }$\varrho=\varrho_{0}$, (3) \emph{the stability condition
}$S\left[\varrho_{0}+\delta\varrho\right]=\mathcal{S}\left(\left\{ F\right\} ;\mathcal{\left\{ V\right\} }\right)+\epsilon$\emph{
with $F_{j}=F_{j}\left[\varrho_{0}+\delta\varrho\right]$ and $\epsilon=\mathcal{O}\left(\left(\delta\varrho\right)^{2}\right)$
is valid for} \emph{any infinitesimal variation }$\delta\varrho$.
Moreover, the associated density $\varrho_{0}$ belongs to the exponential
family. 

The axiomatic formulation leads to several conclusions relevant for
the theory of statistical mechanics. First of all, the notion of the
microcanonical ensemble as well as the postulate of equal a priori
probability follow immediately. It is sufficient to set $M=0$, so
that because $\mathcal{S}$ cannot depend on average values, the exponential
form of \emph{$\varrho_{0}$ }boil\emph{s }down to the constant value.
This value is determined by the energy $E$, the volume $V$ and the
number of particles $N$ which are all the externally fixed variables.
The canonical ensemble appears if we set $M=1$ and take $\tilde{F}_{1}$
to be the Hamiltonian. The exact form of the Hamiltonian (as long
as mathematically reasonable) does not affect the validity of this
simple picture.

Further analysis of the axiomatic definition of entropy brings a new
understanding to notions, such as a generalized (when we consider
more averages than the energy) quasi--static thermodynamic transformation,
or a non--equilibrium state. The first concept is described by a situation
when during the time evolution the phenomenological entropy depends
on the fixed set of parameters, and only the values of the particular
parameters can change. A signature of non-equilibrium appears immediately
when the description based on a certain number $M$ becomes physically
insufficient, so that we need to increase $M$, or the third axiom
is no longer satisfied. From a mathematical point of view, an interesting
question is under which conditions the operation of changing the number
of relevant thermodynamical variables can be done in a continuous,
or even smooth, manner. That could happen by letting the parameters
$\beta_{j}$ related to the new quantities to grow in time, being
identically equal to $0$ in the past. Finally, an interesting perspective
would be to understand if the case $M=\infty$ is a typical scenario
appearing in non--equilibrium statistical mechanics, and if the answer
is yes, to understand how efficiently the system could be described
in terms of a finite number of phenomenologically distinguished parameters.
An adventurous challenge would be to design a kind of $\varrho$-dependent
measure of complexity, able to capture the relevant value of $M$. 

Another recently developing conceptual challenge, namely the attempts
to establish a joined theory of quantum information and quantum thermodynamics
\cite{QIQT1,QIQT2,QIQT3}, could as well benefit from the philosophical
nature of the observation that there is no necessity to maximize the
entropy. In fact, this observation remains valid when other kinds
of accessible information given in terms of non-sharp inequality constraints
on the probability distribution are taken into account. They do not
affect the derivation presented in this letter, but only restrict
the domain of the global variables used. 

Finally, the maximum entropy principle has been extensively used as
a tool to develop new facets of statistical mechanics based on microscopic
entropies different than (\ref{entropy}), eg. (R{\'e}nyi or Tsallis)
\cite{inne1,inne2,inne3,inne4,inne5,inne6}. An important issue would
be to examine these results in the context of the present letter.
The uniqueness property seems to distinguish the logarithmic form
of entropy, the other entropy functionals are thus indeed expected
to go beyond the usual way of reasoning. They might as well turn out
to be unique, provided that additional conditions (axioms) are satisfied.

\acknowledgments

I am indebted to Robert Alicki for several encouraging discussions.
I would like to thank James Lutsko for a valuable correspondence.
Financial support by European Research Council within the project
ODYCQUENT, by the grant number IP2011 046871 of the Polish Ministry
of Science and Higher Education, and the NCN grant number UMO-2012/07/B/ST1/03347
are gratefully acknowledged.

\section*{Appendix}
 The density operator in the case (ii-b)  is given by:
\begin{equation}
\varrho_{0}=\Lambda^{-1}\exp\left(-\sum_{l}\beta_{l}\hat{F}_{l}\right),\quad\Lambda=\textrm{Tr}\,\exp\left(-\sum_{l}\beta_{l}\hat{F}_{l}\right),\label{density-1}
\end{equation}
so that Eq. (11) explicitly reads:

\begin{equation}
F_{j}=f_{j}\left(\beta_{1},\ldots,\beta_{M}\right)=\Lambda^{-1}\textrm{Tr}\left[\hat{F}_{j}\exp\left(-\sum_{l}\beta_{l}\hat{F}_{l}\right)\right].\label{system}
\end{equation}

With the help of the general formula (13) providing the parameter-derivative
of the exponent of a parameter-dependent operator we find:
\begin{equation}
J_{ij}=\frac{\partial f_{j}}{\partial\beta_{i}}=-\left(\textrm{Tr}\left(\hat{F}_{j}\hat{G}_{i}\right)-\textrm{Tr}\left(\hat{F}_{j}\varrho_{0}\right)\textrm{Tr}\hat{G}_{i}\right),
\end{equation}
where
\begin{equation}
\hat{G}_{i}=\Lambda^{-1}\int_{0}^{1}dz\, e^{2\left(1-z\right)\hat{R}}\hat{F}_{i}e^{2z\hat{R}}.\label{G}
\end{equation}

First of all, we observe that since the trace is invariant under cyclic
permutations, we easily get:
\begin{eqnarray}
\textrm{Tr}\hat{G}_{i} & = & \Lambda^{-1}\int_{0}^{1}dz\,\textrm{Tr}\left(e^{2\left(1-z\right)\hat{R}}\hat{F}_{i}e^{2z\hat{R}}\right)\nonumber \\
 & = & \int_{0}^{1}dz\,\textrm{Tr}\left(\hat{F}_{i}\varrho_{0}\right)\nonumber \\
 & = & \textrm{Tr}\left(\hat{F}_{i}\varrho_{0}\right).
\end{eqnarray}
In the second step, we shall split the integration range in (\ref{G})
into two intervals $\left[0,1/2\right]$ and $\left[1/2,1\right]$,
and in the second interval perform the change of variables $z\mapsto1-z$
to get:
\begin{equation}
\hat{G}_{i}=\Lambda^{-1}\!\!\int_{0}^{1/2}\!\!\! dz\left(e^{2\left(1-z\right)\hat{R}}\hat{F}_{i}e^{2z\hat{R}}+e^{2z\hat{R}}\hat{F}_{i}e^{2\left(1-z\right)\hat{R}}\right).
\end{equation}
The above formula in terms of the $\hat{F}_{i}\left(z\right)$ operators
read:
\begin{equation}
\hat{G}_{i}=\int_{0}^{1/2}\!\!\! dz\left(e^{-z\hat{R}}\varrho_{0}\hat{F}_{i}\left(z\right)e^{z\hat{R}}+\textrm{h.c.}\right).
\end{equation}
Using once more the invariance of the trace we thus obtain 
\begin{equation}
\textrm{Tr}\left(\hat{F}_{j}\hat{G}_{i}\right)=\int_{0}^{1/2}\!\!\! dz\textrm{Tr}\left(\hat{F}_{i}\left(z\right)\hat{F}_{j}^{\dagger}\left(z\right)\varrho_{0}+\textrm{h.c.}\right).
\end{equation}
Since the term $\textrm{Tr}\left(\hat{F}_{j}\varrho_{0}\right)\textrm{Tr}\hat{G}_{i}$
does not depend on $z$, it can be ``multiplied'' by $2\int_{0}^{1/2}\!\!\! dz$.
On the other hand, the average values of the dressed operators are
the same as those for the undressed ones, i.e. 
\begin{equation}
\textrm{Tr}\left(\hat{F}_{j}\varrho_{0}\right)=\textrm{Tr}\left(\hat{F}_{j}\left(z\right)\varrho_{0}\right)=\textrm{Tr}\left(\hat{F}_{j}^{\dagger}\left(z\right)\varrho_{0}\right).
\end{equation}
All the above observations boil down to the desired formula {[}Eq.
(14)  and the expression for $J_{ij}$ appearing
above it{]} with the average $\left\langle \cdot\right\rangle _{\varrho_{0}}$
understood in terms of the trace $\textrm{Tr}\left(\cdot\varrho_{0}\right)$.

\end{document}